\documentclass[fleqn,10pt]{wlscirep}

\usepackage[utf8]{inputenc}
\usepackage[T1]{fontenc}
\usepackage{microtype,newpxtext,newpxmath} 
\usepackage{mathtools,gensymb,siunitx,eurosym,url,soul} 
\usepackage[font=footnotesize,textfont=footnotesize]{subcaption}
\usepackage[font=small,textfont=small]{caption}
\usepackage{longtable,array,booktabs,mdwlist}
\usepackage[colorinlistoftodos,prependcaption,textsize=tiny]{todonotes}
\usepackage{comment,todonotes,lineno,acronym}
\usepackage{xcolor,color}
\usepackage{graphicx}

\setkeys{Gin}{draft=false} 
\graphicspath{{figures/} {./}}
\makeatletter
\def\ifdraft{\ifdim\overfullrule>\z@
  \expandafter\@firstoftwo\else\expandafter\@secondoftwo\fi}
\makeatother

\acrodef{ADC}[ADC]{Analog to Digital Converter}
\acrodef{ADM}[ADM]{Asynchronous Delta Modulator}
\acrodef{ADEXP}[AdExp-I\&F]{Adaptive-Exponential Integrate and Fire}
\acrodef{AER}[AER]{Address-Event Representation}
\acrodef{AEX}[AEX]{AER EXtension board}
\acrodef{AE}[AE]{Address-Event}
\acrodef{AFM}[AFM]{Atomic Force Microscope}
\acrodef{AGC}[AGC]{Automatic Gain Control}
\acrodef{AMDA}[AMDA]{AER Motherboard with D/A converters}
\acrodef{ANN}[ANN]{Artificial Neural Network}
\acrodef{API}[API]{Application Programming Interface}
\acrodef{ARM}[ARM]{Advanced RISC Machine}
\acrodef{ASIC}[ASIC]{Application Specific Integrated Circuit}
\acrodef{AdExp}[AdExp-IF]{Adaptive Exponential Integrate-and-Fire}
\acrodef{BCM}[BMC]{Bienenstock-Cooper-Munro}
\acrodef{BD}[BD]{Bundled Data}
\acrodef{BEOL}[BEOL]{Back-end of Line}
\acrodef{BG}[BG]{Bias Generator}
\acrodef{BMI}[BMI]{Brain-Machince Interface}
\acrodef{BTB}[BTB]{band-to-band tunnelling}
\acrodef{CAD}[CAD]{Computer Aided Design}
\acrodef{CAM}[CAM]{Content Addressable Memory}
\acrodef{CAVIAR}[CAVIAR]{Convolution AER Vision Architecture for Real-Time}
\acrodef{CA}[CA]{Cortical Automaton}
\acrodef{CCN}[CCN]{Cooperative and Competitive Network}
\acrodef{CDR}[CDR]{Clock-Data Recovery}
\acrodef{CFC}[CFC]{Current to Frequency Converter}
\acrodef{CHP}[CHP]{Communicating Hardware Processes}
\acrodef{CMIM}[CMIM]{Metal-insulator-metal Capacitor}
\acrodef{CML}[CML]{Current Mode Logic}
\acrodef{CMOL}[CMOL]{Hybrid CMOS nanoelectronic circuits}
\acrodef{CMOS}[CMOS]{Complementary Metal-Oxide-Semiconductor}
\acrodef{CNN}[CCN]{Convolutional Neural Network}
\acrodef{COTS}[COTS]{Commercial Off-The-Shelf}
\acrodef{CPG}[CPG]{Central Pattern Generator}
\acrodef{CPLD}[CPLD]{Complex Programmable Logic Device}
\acrodef{CPU}[CPU]{Central Processing Unit}
\acrodef{CSM}[CSM]{Cortical State Machine}
\acrodef{CSP}[CSP]{Constraint Satisfaction Problem}
\acrodef{CTXCTL}[CTXCTL]{Cortex Control}
\acrodef{CV}[CV]{Coefficient of Variation}
\acrodef{DAC}[DAC]{Digital to Analog Converter}
\acrodef{DAS}[DAS]{Dynamic Auditory Sensor}
\acrodef{DAVIS}[DAVIS]{Dynamic and Active Pixel Vision Sensor}
\acrodef{DBN}[DBN]{Deep Belief Network}
\acrodef{DFA}[DFA]{Deterministic Finite Automaton}
\acrodef{DIBL}[DIBL]{drain-induced-barrier-lowering}
\acrodef{DI}[DI]{delay insensitive}
\acrodef{DMA}[DMA]{Direct Memory Access}
\acrodef{DNF}[DNF]{Dynamic Neural Field}
\acrodef{DNN}[DNN]{Deep Neural Network}
\acrodef{DOF}[DOF]{Degrees of Freedom}
\acrodef{DPE}[DPE]{Dynamic Parameter Estimation}
\acrodef{DPI}[DPI]{Differential Pair Integrator}
\acrodef{DRRZ}[DR-RZ]{Dual-Rail Return-to-Zero}
\acrodef{DRAM}[DRAM]{Dynamic Random Access Memory}
\acrodef{DR}[DR]{Dual Rail}
\acrodef{DSP}[DSP]{Digital Signal Processor}
\acrodef{DVS}[DVS]{Dynamic Vision Sensor}
\acrodef{DYNAP}[DYNAP]{Dynamic Neuromorphic Asynchronous Processor}
\acrodef{EBL}[EBL]{Electron Beam Lithography}
\acrodef{EDVAC}[EDVAC]{Electronic Discrete Variable Automatic Computer}
\acrodef{EEG}[EEG]{electroencephalography}
\acrodef{EIN}[EIN]{Excitatory-Inhibitory Network}
\acrodef{EM}[EM]{Expectation Maximization}
\acrodef{EPSC}[EPSC]{Excitatory Post-Synaptic Current}
\acrodef{EPSP}[EPSP]{Excitatory Post-Synaptic Potential}
\acrodef{EZ}[EZ]{Epileptogenic Zone}
\acrodef{FDSOI}[FDSOI]{Fully-Depleted Silicon on Insulator}
\acrodef{FET}[FET]{Field-Effect Transistor}
\acrodef{FFT}[FFT]{Fast Fourier Transform}
\acrodef{FI}[F-I]{Frequency-Current}
\acrodef{FPGA}[FPGA]{Field Programmable Gate Array}
\acrodef{FR}[FR]{Fast Ripple}
\acrodef{FSA}[FSA]{Finite State Automaton}
\acrodef{FSM}[FSM]{Finite State Machine}
\acrodef{GIDL}[GIDL]{gate-induced-drain-leakage}
\acrodef{GOPS}[GOPS]{Giga-Operations per Second}
\acrodef{GPU}[GPU]{Graphical Processing Unit}
\acrodef{GUI}[GUI]{Graphical User Interface}
\acrodef{HAL}[HAL]{Hardware Abstraction Layer}
\acrodef{HFO}[HFO]{High Frequency Oscillation}
\acrodef{HH}[H\&H]{Hodgkin \& Huxley}
\acrodef{HMM}[HMM]{Hidden Markov Model}
\acrodef{HRS}[HRS]{High-Resistive State}
\acrodef{HR}[HR]{Human Readable}
\acrodef{HSE}[HSE]{Handshaking Expansion}
\acrodef{HW}[HW]{Hardware}
\acrodef{ICT}[ICT]{Information and Communication Technology}
\acrodef{IC}[IC]{Integrated Circuit}
\acrodef{IEEG}[iEEG]{intracranial electroencephalography}
\acrodef{IF2DWTA}[IF2DWTA]{Integrate \& Fire 2--Dimensional WTA}
\acrodef{IFSLWTA}[IFSLWTA]{Integrate \& Fire Stop Learning WTA}
\acrodef{IF}[I\&F]{Integrate-and-Fire}
\acrodef{IMU}[IMU]{Inertial Measurement Unit}
\acrodef{INCF}[INCF]{International Neuroinformatics Coordinating Facility}
\acrodef{INI}[INI]{Institute of Neuroinformatics}
\acrodef{IO}[I/O]{Input/Output}
\acrodef{IPSC}[IPSC]{Inhibitory Post-Synaptic Current}
\acrodef{IPSP}[IPSP]{Inhibitory Post-Synaptic Potential}
\acrodef{IP}[IP]{Intellectual Property}
\acrodef{ISI}[ISI]{Inter-Spike Interval}
\acrodef{IoT}[IoT]{Internet of Things}
\acrodef{JFLAP}[JFLAP]{Java - Formal Languages and Automata Package}
\acrodef{LEDR}[LEDR]{Level-Encoded Dual-Rail}
\acrodef{LFP}[LFP]{Local Field Potential}
\acrodef{LLC}[LLC]{Low Leakage Cell}
\acrodef{LNA}[LNA]{Low-Noise Amplifier}
\acrodef{LPF}[LPF]{Low Pass Filter}
\acrodef{LRS}[LRS]{Low-Resistive State}
\acrodef{LSM}[LSM]{Liquid State Machine}
\acrodef{LTD}[LTD]{Long Term Depression}
\acrodef{LTI}[LTI]{Linear Time-Invariant}
\acrodef{LTP}[LTP]{Long Term Potentiation}
\acrodef{LTU}[LTU]{Linear Threshold Unit}
\acrodef{LUT}[LUT]{Look-Up Table}
\acrodef{LVDS}[LVDS]{Low Voltage Differential Signaling}
\acrodef{MCMC}[MCMC]{Markov-Chain Monte Carlo}
\acrodef{MEMS}[MEMS]{Micro Electro Mechanical System}
\acrodef{MFR}[MFR]{Mean Firing Rate}
\acrodef{MIM}[MIM]{Metal Insulator Metal}
\acrodef{MLP}[MLP]{Multilayer Perceptron}
\acrodef{MOSCAP}[MOSCAP]{Metal Oxide Semiconductor Capacitor}
\acrodef{MOSFET}[MOSFET]{Metal Oxide Semiconductor Field-Effect Transistor}
\acrodef{MOS}[MOS]{Metal Oxide Semiconductor}
\acrodef{MRI}[MRI]{Magnetic Resonance Imaging}
\acrodef{NDFSM}[NDFSM]{Non-deterministic Finite State Machine} 
\acrodef{ND}[ND]{Noise-Driven}
\acrodef{NEF}[NEF]{Neural Engineering Framework}
\acrodef{NHML}[NHML]{Neuromorphic Hardware Mark-up Language}
\acrodef{NIL}[NIL]{Nano-Imprint Lithography}
\acrodef{NMDA}[NMDA]{N-Methyl-D-Aspartate}
\acrodef{NME}[NE]{Neuromorphic Engineering}
\acrodef{NN}[NN]{Neural Network}
\acrodef{NRZ}[NRZ]{Non-Return-to-Zero}
\acrodef{NSM}[NSM]{Neural State Machine}
\acrodef{OR}[OR]{Operating Room}
\acrodef{OTA}[OTA]{Operational Transconductance Amplifier}
\acrodef{PCB}[PCB]{Printed Circuit Board}
\acrodef{PCHB}[PCHB]{Pre-Charge Half-Buffer}
\acrodef{PCM}[PCM]{Phase Change Memory}
\acrodef{PE}[PE]{Phase Encoding}
\acrodef{PFA}[PFA]{Probabilistic Finite Automaton}
\acrodef{PFC}[PFC]{prefrontal cortex}
\acrodef{PFM}[PFM]{Pulse Frequency Modulation}
\acrodef{PR}[PR]{Production Rule}
\acrodef{PSC}[PSC]{Post-Synaptic Current}
\acrodef{PSP}[PSP]{Post-Synaptic Potential}
\acrodef{PSTH}[PSTH]{Peri-Stimulus Time Histogram}
\acrodef{QDI}[QDI]{Quasi Delay Insensitive}
\acrodef{RA}[RA]{Resected Area}
\acrodef{RAM}[RAM]{Random Access Memory}
\acrodef{RDF}[RDF]{random dopant fluctuation}
\acrodef{RELU}[ReLu]{Rectified Linear Unit}
\acrodef{RLS}[RLS]{Recursive Least-Squares}
\acrodef{RMSE}[RMSE]{Root Mean Squared-Error}
\acrodef{RMS}[RMS]{Root Mean Squared}
\acrodef{RNN}[RNN]{Recurrent Neural Networks}
\acrodef{ROLLS}[ROLLS]{Reconfigurable On-Line Learning Spiking}
\acrodef{RRAM}[R-RAM]{Resistive Random Access Memory}
\acrodef{R}[R]{Ripples}
\acrodef{SAC}[SAC]{Selective Attention Chip}
\acrodef{SAT}[SAT]{Boolean Satisfiability Problem}
\acrodef{SCX}[SCX]{Silicon CorteX}
\acrodef{SD}[SD]{Signal-Driven}
\acrodef{SEM}[SEM]{Spike-based Expectation Maximization}
\acrodef{SLAM}[SLAM]{Simultaneous Localization and Mapping}
\acrodef{SNN}[SNN]{Spiking Neural Network}
\acrodef{SNR}[SNR]{Signal to Noise Ratio}
\acrodef{SOC}[SOC]{System-On-Chip}
\acrodef{SOI}[SOI]{Silicon on Insulator}
\acrodef{SOZ}[SOZ]{Seizure Onset Zone}
\acrodef{SP}[SP]{Separation Property}
\acrodef{SRAM}[SRAM]{Static Random Access Memory}
\acrodef{STDP}[STDP]{Spike-Timing Dependent Plasticity}
\acrodef{STD}[STD]{Short-Term Depression}
\acrodef{STP}[STP]{Short-Term Plasticity}
\acrodef{STT-MRAM}[STT-MRAM]{Spin-Transfer Torque Magnetic Random Access Memory}
\acrodef{STT}[STT]{Spin-Transfer Torque}
\acrodef{SW}[SW]{Software}
\acrodef{TCAM}[TCAM]{Ternary Content-Addressable Memory}
\acrodef{TLE}[TLE]{Temporal Lobe Epilepsy}
\acrodef{TFT}[TFT]{Thin Film Transistor}
\acrodef{USB}[USB]{Universal Serial Bus}
\acrodef{VHDL}[VHDL]{VHSIC Hardware Description Language}
\acrodef{VLSI}[VLSI]{Very Large Scale Integration}
\acrodef{VOR}[VOR]{Vestibulo-Ocular Reflex}
\acrodef{WCST}[WCST]{Wisconsin Card Sorting Test}
\acrodef{WTA}[WTA]{Winner-Take-All}
\acrodef{XML}[XML]{eXtensible Mark-up Language}
\acrodef{CTXCTL}[CTXCTL]{CortexControl}
\acrodef{divmod3}[DIVMOD3]{divisibility of a number by three}
\acrodef{hWTA}[hWTA]{hard Winner-Take-All}
\acrodef{sWTA}[sWTA]{soft Winner-Take-All}

\title{An electronic neuromorphic system for real-time detection of High Frequency Oscillations (HFOs) in intracranial EEG}

\author[1,2*]{Mohammadali Sharifshazileh}
\author[1,2*]{Karla Burelo}
\author[2]{Johannes Sarnthein}
\author[1]{Giacomo Indiveri}
\affil[1]{Institute of Neuroinformatics, University of Zurich and ETH Zurich,  Zurich, 8057, Switzerland}
\affil[2]{Klinik für Neurochirurgie, UniversitätsSpital und Universität Zürich,  8091 Zurich, Switzerland}
\affil[*]{These authors contributed equally.}

\begin{abstract}

  The analysis of biomedical signals for clinical studies and therapeutic applications can benefit from compact and portable devices that can process these signals locally, in real-time, without the need for off-line processing. An example is the recording of intracranial EEG (iEEG) during epilepsy surgery with the detection of High Frequency Oscillations (HFOs, 80-500 Hz), which are a biomarker for the epileptogenic zone.

  Conventional approaches of HFO detection involve the offline analysis of prerecorded data, often on bulky computers. However, clinical applications during surgery or in long-term intracranial recordings demand a self-sufficient embedded device that is battery-powered to avoid interfering with other electronic equipment in the operation room.

  Mixed-signal and analog-digital neuromorphic circuits offer the possibility of building compact, embedded and low-power neural network processing systems that can analyze data on-line and produce results with short latency in real-time. These characteristics are well suited for clinical applications that involve the processing of biomedical signals at (or very close to) the sensor level.

  In this work, we present a neuromorphic system that combines for the first time a neural recording headstage with a signal-to-spike conversion circuit and a multi-core spiking neural network (SNN) architecture on the same die
  for recording, processing, and detecting clinically relevant HFOs in iEEG from epilepsy patients.
  
  The device  was fabricated using a standard 0.18\,$\mu$m CMOS technology node and has a total area of 99\,mm$^{2}$. We demonstrate its application to HFO detection in the iEEG recorded from 9 patients with temporal lobe epilepsy who subsequently underwent epilepsy surgery. The total average power consumption of the chip during the detection task was 614.3\,$\mu$W. We show how the neuromorphic system  can reliably detect HFOs: the system predicts postsurgical seizure outcome with state-of-the-art accuracy, specificity and sensitivity (78\%, 100\%, and 33\% respectively).

  This is the first feasibility study towards identifying relevant features in intracranial human data in real-time, on-chip, using event-based 
  processors and spiking neural networks. By providing ``neuromorphic intelligence'' to neural recording circuits the approach proposed will pave the way for the development of systems that can detect HFO areas directly in the operation room and improve the seizure outcome of epilepsy surgery.

\end{abstract}

\begin{document}

\flushbottom
\maketitle
%
%
\newpage

\section{Introduction}
\label{sec:introduction}

\ifdraft{\subsection*{Background - setting the stage}}

The amount and type of sensory data that can be recorded is continuously increasing due to the recent progress in microelectronic technology~\cite{NatureBigData18}. This data deluge calls for the development of low-power embedded edge computing technologies that can process the signals being measured locally, without requiring bulky computers or the need for internet connection and cloud servers. In particular, biomedical signal processing and clinical applications will greatly benefit from a direct and local processing of physiological measurements using compact and portable devices.
Spiking neural networks (SNNs) have been proposed as a powerful computing paradigm for processing spatio-temporal signals and detecting complex patterns within them~\cite{Maass_Sontag00,Kasabov_etal13}, so they are well suited for biomedical applications.

\ifdraft{\subsection*{Novel edge-computing technologies}}

Neuromorphic engineering~\cite{Indiveri_Horiuchi11} has produced mixed signal neuromorphic circuits that can directly emulate the physics of real neurons to implement faithful models of neural dynamics and SNNs~\cite{Mead90,Chicca_etal14}.
By integrating many silicon neurons onto microelectronic chips, it is possible to build compact and low-power SNN processing systems to efficiently process time-varying signals in real time.
\ifdraft{\subsection*{Previous work - state-of-the-art and open challenges}}
Examples of neuromorphic systems developed with this goal in mind have already been applied to processing Electrocardiogram (ECG) or Electromyography (EMG) signals~\cite{Bauer_etal19,Corradi_etal19,Donati_etal19,Azghadi_etal20}.

However, these systems were sub-optimal, as they required external bio-signal recording, frontend devices, and data conversion interfaces. Bio-signal recording headstages typically comprise analog circuits to amplify and filter the signals being measured, and can be highly diverse in specifications depending on the application~\cite{Harrison08}. For example, neural recording headstages for experimental neuroscience target high-density recordings~\cite{Jun_etal17,Frey_etal10,Ballini_etal14,Khazaei_etal20} and minimize the circuit area requirements, while devices used for clinical studies and therapeutic applications require a small number of recording channels and the highest possible signal-to-noise ratio (SNR)~\cite{Boran_etal19, Zijlmans_etal17, Fedele_etal17a, Mohammadi_etal14}

\ifdraft{\subsection*{Novelty of the proposed approach}}

In this work, we present a neuromorphic system that combines for the first time a neural recording headstage with a signal-to-spike conversion circuit and a multi-core SNN architecture on the same die for recording, processing, and detecting clinically relevant biomarkers in intracranial EEG recordings (iEEG) from epilepsy patients.

\ifdraft{\subsection*{Specific problem addressed}}

Epilepsy is the most common severe neurological disease. In about one-third of patients, seizures cannot be controlled by medication. Selected patients with focal epilepsy can achieve seizure freedom if the epileptogenic zone (EZ), which is the brain area generating the seizures, is correctly identified and surgically removed in its entirety. Presurgical and intraoperative measurement of iEEG signals is often needed to identify the EZ precisely~\cite{Zijlmans_etal19, Lesko_etal20, Jobst_etal20,Farahmand_etal19}.
High Frequency Oscillations (HFOs) have been proposed as a new biomarker in iEEG to delineate the EZ~\cite{Fedele_etal17b, Zijlmans_etal17, Frauscher_etal17, Demuru_etal20, Vant-klooster_etal17, Fedele_etal16, Fedele_etal17a}
While HFOs have been historically divided into “ripples” (80–250 Hz) and “fast ripples” (FR, 250–500 Hz), detection of their co-occurrence was shown to enable the optimal prediction of postsurgical seizure freedom~\cite{Fedele_etal17b}. In that study, HFOs were detected automatically by a software algorithm that matched the morphology of the HFO to a predefined template (Morphology Detector)~\cite{Fedele_etal17b, Burnos_etal16}
An example of such an HFO is shown in Fig.~\ref{fig:pipeline_combined}a.
While software algorithms are used for detecting HFOs offline, compact embedded neuromorphic systems that can record iEEGs and detect HFOs online, in real time, would be able to provide valuable information during surgery, and simplify the collection of statistics in long-term epilepsy monitoring\cite{Beniczky_etal20, Stirling_etal20}.
Here we show how our neuromorphic system paves the way toward the development of a compact embedded battery-operated system that can be used for the real-time online detection of HFOs.

\ifdraft{\subsection*{Outline of the manuscript}}

In this paper we first describe the design principles of the HFO detection architecture and its neuromorphic circuit implementation. We discuss the characteristics of the circuit blocks proposed and present experimental results measured from the fabricated device. We then show how the neuromorphic system performs in HFO detection compared to the Morphology Detector~\cite{Fedele_etal17b} on iEEG recorded from the medial temporal lobe~\cite{Fedele_etal17c}\footnote{
A short video describing the rationale underlying the study is available on the \href{https://www.bci-award.com/Home}{Annual BCI award 2020 webpage} at: \url{https://youtu.be/Pw83Mrza_rg}}

\begin{figure}
\centering
\includegraphics[width=\linewidth]{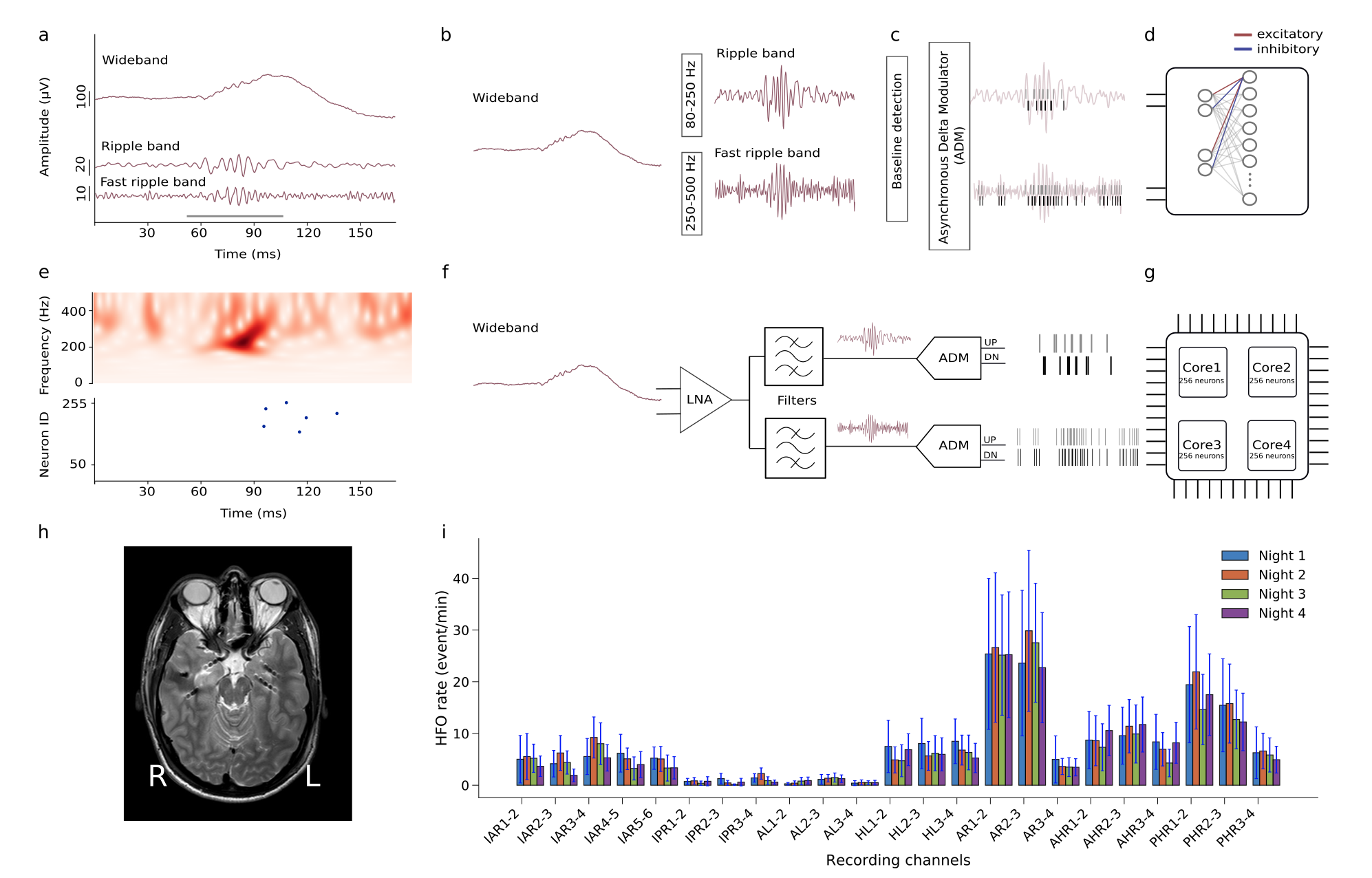}
\caption{Automatic HFO detection using a bio-inspired SNN. (a) The pre-recorded iEEG signal in wideband, Ripple band (80-250Hz) and Fast Ripple band (250-500Hz). HFOs stand out of the baseline in the signal. The period marked by the gray bar represents a clinically relevant HFO~\cite{Fedele_etal17b,Fedele_etal17c}. (b,c,d) Software simulated spiking neural network (SNN). For preprocessing the wideband EEG is filtered in Ripple band and Fast Ripple band. A baseline detection stage finds the optimum threshold that is applied in an Asynchronous Delta Modulator (ADM) which converts the signal to spikes. Signal traces are encoded by UP spikes (gray bars) and DOWN spikes (black bars), which are then fed as input into the SNN. The SNN is implemented as a two-layer spiking network of integrate and fire neurons with dynamic synapses. Each neuron in the second layer receives four inputs: two excitatory spike trains from UP channels and two inhibitory ones from DOWN channels. The parameters of the network were chosen to exhibit the relevant temporal dynamics and tune the neurons to produce output spikes in response to input spike train patterns that encode clinically relevant HFOs. (e, top) Time-frequency spectrum of the Fast Ripple iEEG of panel a. (e, bottom) Firing of SNN neurons indicate the occurrence of the HFO. 
(f) Block diagram of the neuromorphic system input headstage. The headstage comprises a low noise amplifier (LNA), two configurable bandpass filters and two ADM circuits that convert the analog waveforms into spike trains. 
(g) The spikes produced by the ADMs are sent to a multi-core array of silicon neurons that are configured to implement the desired SNN. (h) MRI with 7 implanted depth electrodes that sample the mesial temporal structures of a patient with drug resistant temporal lobe epilepsy (Patient 1). 
(i) Rates of HFOs detected by the neuromorphic SNN for recordings made across four nights for Patient 1. HFO rate and variability across intervals within a night is indicated by standard error bars. Recording channels AR1-2 and AR2-3 in right amygdala showed the highest HFO rates which were stable over nights. Thus, the neuromorphic system would predict that a therapeutic resection, which should include the right amygdala, would achieve seizure freedom. Indeed, a resection including the right amygdala achieved seizure freedom for >1 year.}
\label{fig:pipeline_combined}
\end{figure}

\section{Results}
\label{sec:results}

Figure~\ref{fig:pipeline_combined} shows how prerecorded iEEG~\cite{Fedele_etal17c} was processed by the frontend headstage and the SNN multi-core neuromorphic architecture. Signals were band-passed filtered into Ripple and Fast Ripple bands (Fig.~\ref{fig:pipeline_combined}a,b,f). The resulting waveforms were converted into spikes using asynchronous delta modulator (ADM) circuits~\cite{Yang_etal15,Corradi_Indiveri15} (Fig.~\ref{fig:pipeline_combined}c,f) and fed into the SNN architecture (Fig.~\ref{fig:pipeline_combined}d,g).
Neuronal spiking signalled the detection of an HFO (Fig.~\ref{fig:pipeline_combined}e bottom).
All stages were first simulated in software to find optimal parameters and then validated with the hardware components. 
The HFO detection was validated by comparing the HFO rate across recording intervals (Fig.~\ref{fig:pipeline_combined}i) and against postsurgical seizure outcome~\cite{Fedele_etal17b}.

\begin{figure}
 \centering\includegraphics[width = \linewidth]{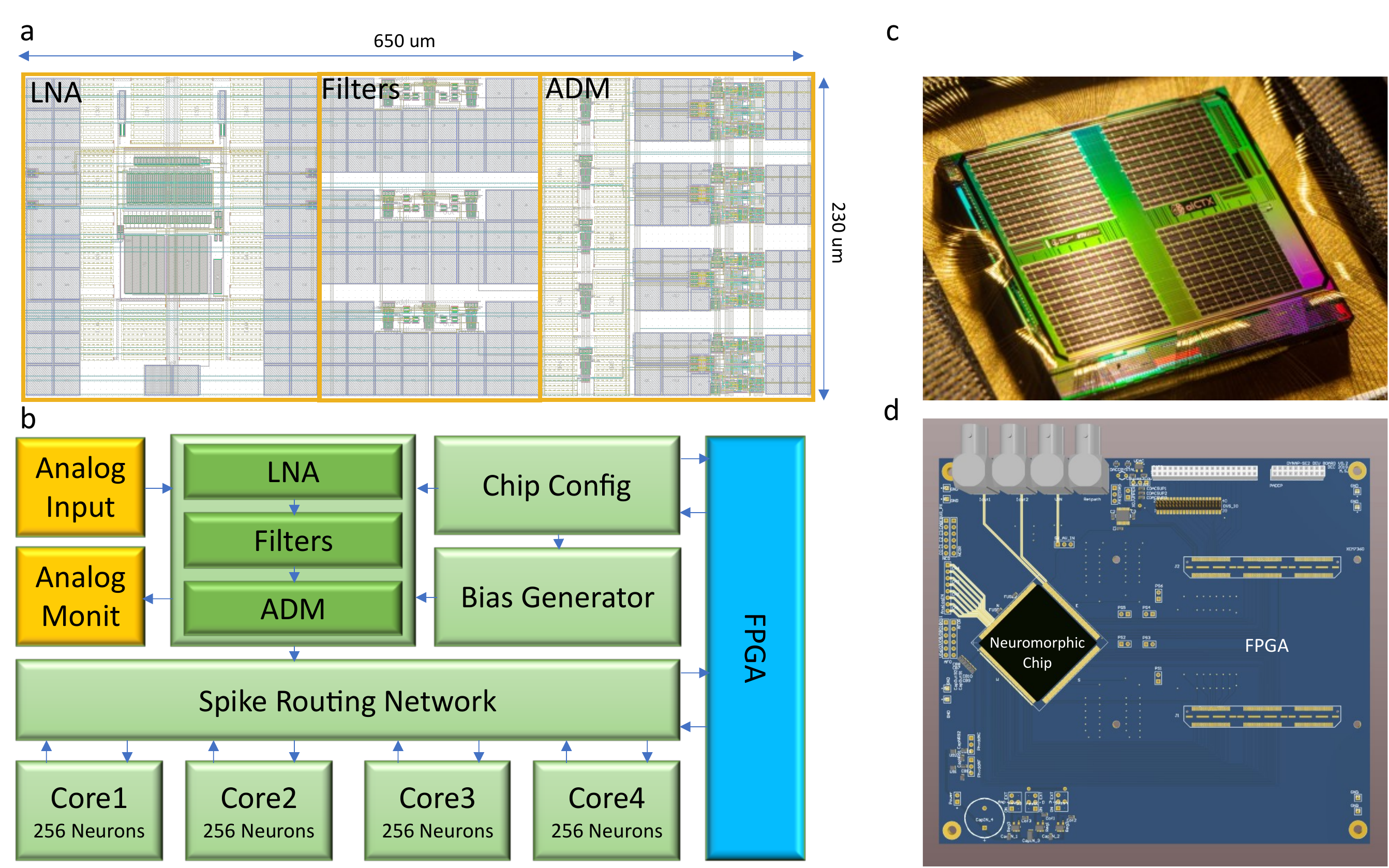}
 \caption{Neuromorphic electronic system overview (a) Physical layout of a single channel of the analog headstage array, including the LNA, three low-pass/band-pass filters, and four ADM signal to spike encoders. (b) Reduced block diagram of the neuromorphic platform. Analog signals from electrodes are fed into the input headstage that converts them into spike trains and sends them to the SNN implemented on the multi-neuron cores, via a spike routing network. The spike routing network routes the spikes within on-chip SNN and to an external FPGA device used for data logging and prototyping. The FPGA is also used for setting circuit parameters (c) Chip photograph showing the 11\,mm x 9\,mm silicon die. (d) Prototyping Printed Circuit Board (PCB) used to host the chip and the infrastructure to implement the test setup. The setup is composed of a prototyping FPGA board mounted on the same PCB that hosts the chip, and of probe points to evaluate the characteristics of both input headstage and SNN multi-core network.}
 \label{fig:hw-overview}
\end{figure}

\subsection{The neuromorphic system}

An overview of the hardware neuromorphic system components is depicted in Fig.~\ref{fig:hw-overview}. The chip (Fig~\ref{fig:hw-overview}c) was fabricated using a standard 180\,nm \ac{CMOS} process. It comprises 8 input channels (headstages) responsible for the neural recording operation, band-pass filtering and conversion to spikes~\cite{Sharifshazileh_etal19}, and a multi-core neuromorphic processor with 4 neurosynaptic cores of 256 neurons each, which is a novel type of Dynamic Neuromorphic Asynchronous Processor (DYNAP) based on the DYNAP-SE device~\cite{Moradi_etal18}. 
The total chip area is 99\,mm$^{2}$. The 8 headstages occupy 1.42\,mm$^{2}$ with a single headstage occupying an area of 0.15\,mm$^{2}$ (see Fig.~\ref{fig:hw-overview}a). The area of the four SNN cores is 77.2\,mm$^{2}$ with a single SNN core occupying 15\,mm$^{2}$.
For the HFO detection task, the total average power consumption of the chip, at the standard 1.8\,V supply voltage, was 614.3\,$\mu$W.  The total static power consumption of a single headstage was 7.3\,$\mu$W. The conversion of filtered waveforms to spikes by the ADMs consumed on average 109.17\,$\mu$W. The power required by the SNN synaptic circuits to process the spike rates produced by the ADMs was 497.82\,$\mu$W, while the power required by  the neurons in the second layer of the SNN to produce the output spike rates was 0.2\,nW.
The block diagram of the hardware system functional modules is shown in Fig.~\ref{fig:hw-overview}b. The \ac{FPGA} block the right of the figure represents a prototyping device that is used only for characterizing the system performance. Figure~\ref{fig:hw-overview}c shows the chip photograph, and Fig.~\ref{fig:hw-overview}d represents a rendering of the prototyping \ac{PCB} used to host the chip.

\begin{figure}
\centering
\includegraphics[width=1\linewidth]{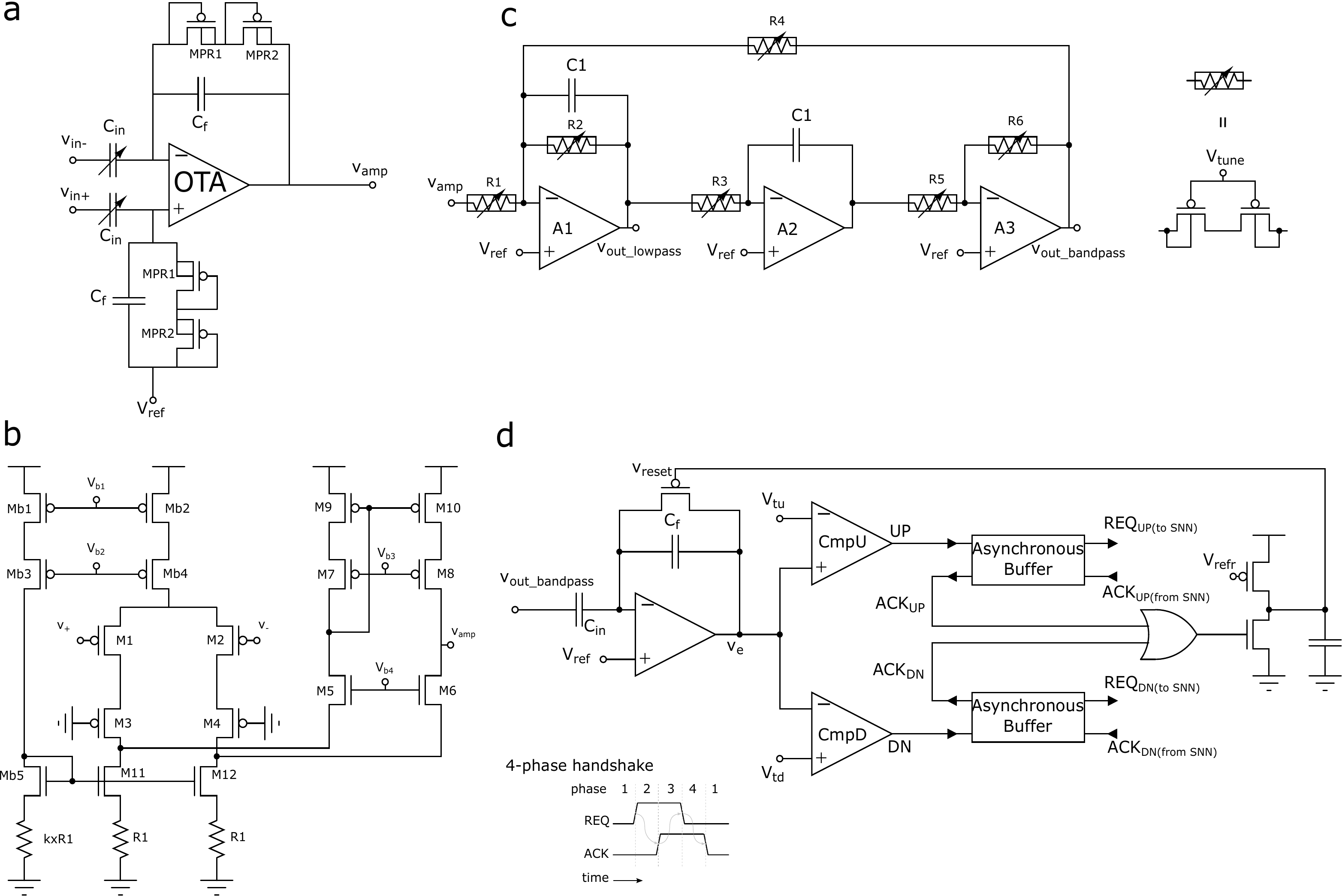}
\caption{Schematic diagrams of the input headstage circuits. (a) Variable-gain LNA using variable input capacitor array and pseudo-resistors. The gain of this stage is calculated by  $C_{in}$/$C_{f}$; the use of the pseudo-resistors allows to reach small low cut-off frequencies. (b)  Folded cascode OTA using resistive degeneration to reduce the noise influence of nMOS devices. Note that the current flowing through the bias branch is $k$ times smaller than the tail branch of the amplifier. (c) band-pass (Tow-Thomas) filters for performing second-order filtering in both ripple and fast-ripple bands as well as low-frequency component of the iEEG. Tunable pseudo resistors are used to adjust the filter gain, center-frequency and band-width. The same basic structure can be used to provide both low-pass and band-pass outputs, thus is desirable in terms of design flexibility. (d) \acf{ADM} circuit to convert the analog filter outputs into spike trains. The \ac{ADM} input amplifier has a gain of $C_{in}$/$C_{f}$ in normal operation when $V_{reset}$ is low and the feedback PMOS switch is off. As the amplified signal crosses one of the two  thresholds, $V_{tu}$ or $V_{td}$, an UP or DN spike is produced by asserting the corresponding REQ signal. A 4-phase handshaking mechanism produces the corresponding ACK signal in response to the spike. Upon receiving the ACK signal, the \ac{ADM} resets the amplifier input and goes back to normal operation after a refractory period determined by the value of $V_{refr}$. The asynchronous buffers act as 4-phase handshaking interfaces that propagate the UP/DN signals to the on-chip \ac{AER} spike routing network of Fig.~\ref{fig:hw-overview}.}
\label{fig:schematic_AF}
\end{figure}

\ifdraft{\subsubsection*{The Low-Noise Amplifier (Fig.3a)}}

Figure~\ref{fig:schematic_AF} shows the details of the main circuits used in a single channel of the input headstage.
In particular, the schematic diagram of the \ac{LNA} is shown in Fig.~\ref{fig:schematic_AF}a. It consists of an Operational Transconductance Amplifier (OTA) with variable input \ac{MIM} capacitors, $C_{in}$ that can be set to 2/8/14/20\,pF and a Resistor-Capacitor (RC) feedback in which the resistive elements are implemented using MOS-bipolar structures~\cite{Harrison_Charles03}.
The MOS-bipolar pseudoresistors $MPR1$ and $MPR2$, and the capacitors $C_f$=200\,fF of Fig.~\ref{fig:schematic_AF}a were chosen to implement a high-pass filter with a low cutoff frequency of 0.9\,Hz. Similarly, the input capacitors $C_{in}$, $C_f$, and the transistors of the OTA were sized to produce maximum amplifier gain of approximately 40.2\,dB but can be adjusted to smaller values by changing the capacitance of $C_{in}$. 

\ifdraft{\subsubsection*{The Operational Transconductance Amplifier (Fig.3b)}}

Figure~\ref{fig:schematic_AF}b shows the schematic of the OTA, which is a modified version of a standard folded-cascode topology~\cite{Wattanapanitch_etal07}. The currents of the transistors in the folded branch (M5-M10) are scaled to approximately 1/6th of the currents in the original branch M1-M4. The noise generated by M5-M10 is negligible compare to that of M1-M4 due to the low current in these transistors. As a result, the total current and the total input-referred noise of the OTA was minimized.

To ensure accurate bias-current scaling, the currents of Mb2 and Mb4 in Fig.~\ref{fig:schematic_AF}b were set using the bias circuit formed by Mb1, Mb3, and Mb5. The voltages $Vb1$ and $Vb2$ in the biasing circuit can be set by a programmable 6526-level integrated bias-generator, integrated on chip~\cite{Delbruck_Van-Schaik05}. The current sources formed by Mb1 and Mb2 were cascoded to increase their output impedance and to ensure accurate current scaling. These devices operate in strong inversion to reduce the effect of threshold voltage variations. The source-degenerated current mirrors formed by M11, M12, Mb5 and resistors R1 and $k\times$R1 assure that the currents in M5 and M6 are a small fraction of the currents in M3 and M4. The R1 gain coefficient $k$ was chosen at design time to be $k=8.5$. 
Thanks to the use of this source-degenerated current source scheme, the $1/f$ noise in the OTA is limited mainly to the effect of  the input differential pair. Therefore, the transistors of the input-differential pair were chosen to be pMOS devices and to have a large area.

\ifdraft{\subsubsection*{The band-pass filters (Fig.3c)}}

The active filters implemented in our system are depicted in Fig.~\ref{fig:schematic_AF}c. They comprise three operational amplifiers, configured to form a Tow-Thomas resonating architecture~\cite{Fleischer_Tow73}. This architecture consists of a damped inverting integrator that is stabilized by $R2$ and cascaded with another undamped integrator, and an inverting amplifier for adjusting the loop-gain by a factor set by the ratio $R6/R5$. The center frequency $f_{0}$ of the bandpass filter can be calculated as  $f_{0}={1}/{2\pi \sqrt{R3 R4 C1^{2}}}$. By choosing $R3=R4=R$, we can then simplify it to $f_{0}={1}/{2\pi RC1}$. Similarly, the gain of the filter is $\left | T_{BP} \right |={R4}/{R1}$ and its bandwidth $BW={2\pi f_{0}\sqrt{R3 R4}}/{R2}$, but with our choice of resistors we can show that $\left | T_{BP} \right |=R/R1$ and $BW=1/{(R2 C1)}$. 
Therefore, this analysis shows that  $R$ is responsible for setting $f_{0}$, $R1$ for adjusting the gain, and $R2$ for tuning the bandwidth. Moreover, due to the resistive range of the tunable double-PMOS pseudo resistors used in this design~\cite{Shiue_etal15},  $f_{0}$ was set in the sub-hundred Hertz region by choosing $C1=10\,pF$.

\ifdraft{\subsubsection*{The Asynchronous Delta Modulator (Fig.3d)}}

Figure~\ref{fig:schematic_AF}d shows the schematic diagram of the ADM circuit~\cite{Corradi_Indiveri15}. There are four of such circuits per headstage channel. One for converting the wideband signal $V_{amp}$ into spike trains; one for converting the output of the low-pass filter $V_{out\_lowpass}$; and two for converting the output of the two band-pass filters $V_{out\_bandpass}$. 
The amplifier at the input of the ADM circuit in Fig.~\ref{fig:schematic_AF}d implements an adaptive feedback amplification stage with a gain set by $C_{in}/C_{f}$ that in our design is equal to 8 when $V_{reset}$ is high, and approximately zero during periods in which $V_{reset}$ is low. In these periods, defined as ``reset assertion'' the output of the amplifier $V_{e}$ is clamped to $V_{ref}$, while in periods when $V_{reset}$ is high, the output voltage $V_{e}$ represents the amplified version of the input. 
The $V_{e}$  signal is then sent as input to a pair of comparators that produce either ``UP'' or ``DN'' digital pulses depending if $V_{e}$ is greater than $V_{tu}$ or lower than $V_{td}$. These parameters set the ADM circuit sensitivity to the amplitude of the Delta-change. The smallest values that these voltages can take is limited by the input offset of the ADM comparators (see $CmpU, CmpD$ in Fig.~\ref{fig:schematic_AF}d), which is approximately 500\,$\mu$V.

Functionally, the ADM represents a Delta-modulator that quantizes the difference between the current amplitude of $V_{e}$ and the amplitude of $V_{e}$  at the previous reset assertion. The precise timing of the $UP/DN$ spikes produced in this way are deemed to contain all the information about the original input signal, given that the parameters of the ADM are known~\cite{Lazar_etal13}.
The UP and DN spikes are used as the request signals of the asynchronous \ac{AER} communication protocol~\cite{Mahowald92a,Deiss_etal98,Boahen98} used by the spike routing network for transmitting the spikes to the silicon neurons of the neuromorphic cores (Fig.~\ref{fig:hw-overview}).
We call this event-based computation.
These signals are pipelined through  asynchronous buffers that locally generates $ACK_{UP/DN}$ to reset the ADM with every occurrence of an UP or DN event.
The output of the asynchronous buffer $REQ_{UP/DN(to SNN)}$  conveys these events to the next asynchronous stages.
The bias voltage $V_{refr}$ controls the refractory period that keeps the amplifier reset and limits the maximum event rate of the circuit to reduce power consumption. The bias voltages $V_{tu}$ and $V_{td}$ control the sensitivity of the ADM and the number of spikes produced per second, with smaller values producing spike trains with higher frequencies. Small $V_{tu}$ and $V_{td}$ settings lead to higher power consumption and allow the faithful reconstruction of the input signal with all its frequency components. The ADM hyper parameters $V_{refr}, V_{tu}$, and $V_{td}$ can therefore be optimized to achieve high reconstruction accuracy of the input signal and suppress background noise (e.g., due to high-frequency signal components), depending on the nature of the signal being processed (see Methods).
All of the 32 ADM output channels are then connected to a common AER encoder which includes an AER arbiter~\cite{Boahen00} used to manage the asynchronous traffic of events and convey them to the on-chip spike routing network.

\subsection{Analog headstage circuit measurements}

\begin{figure}
\centering
\includegraphics[width=1\linewidth]{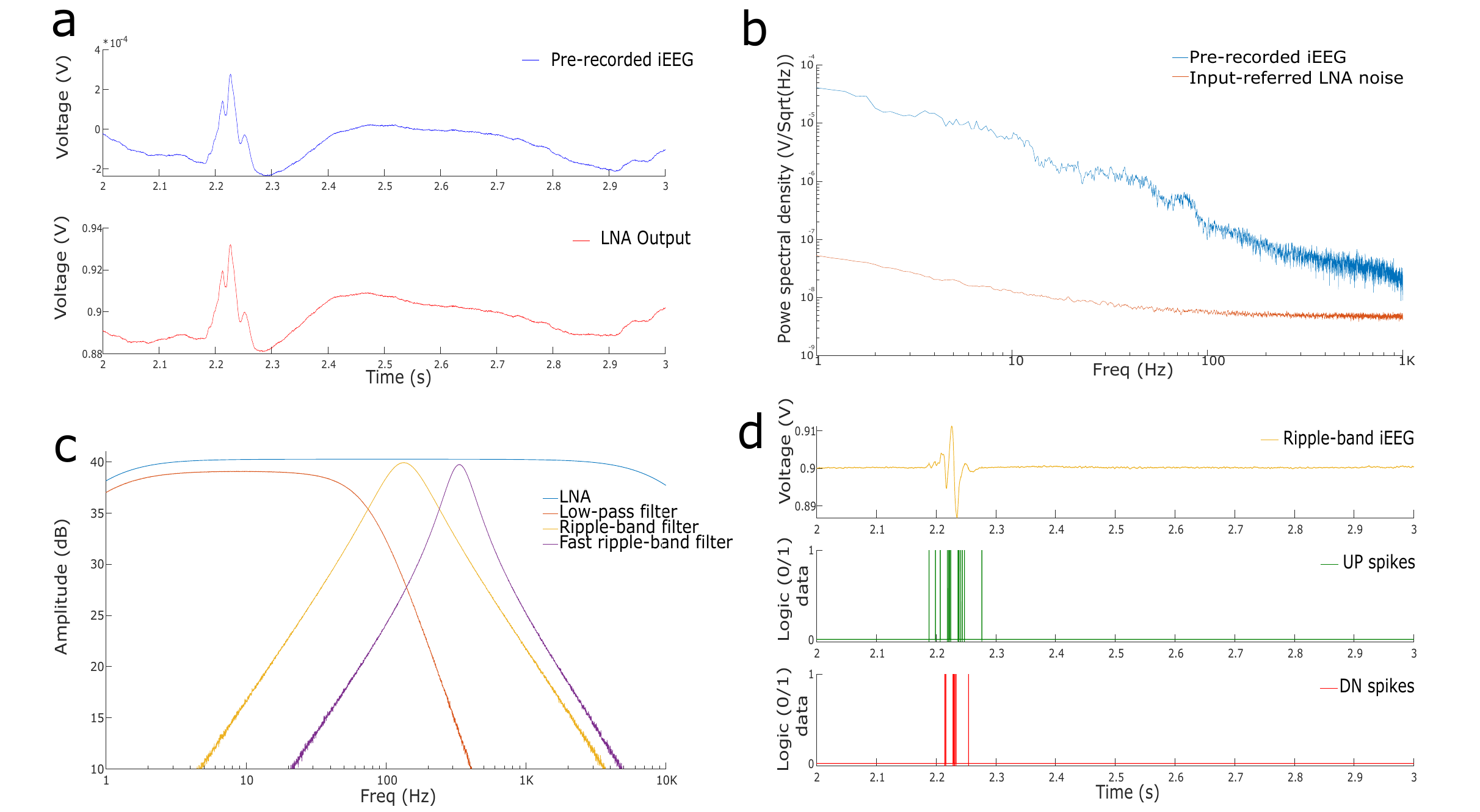}
\caption{Measurements from the analog headstage. (a) An iEEG sample~\cite{Fedele_etal17c} and the LNA amplified output. (b) Noise floor of the headstage LNA and iEEG power spectral density. In the HFO range (80-250 Hz) the noise level of the LNA is below the iEEG noise floor by an order of magnitude. (c) Frequency response of the implemented filters in the headstage. The band-pass filters are tuned to highlight HFO frequency bands. (d) ADM output spikes in response to the ripple-filtered signal. The top plot shows the analog filter output, the middle plot shows the UP spikes generated by the ADM and the bottom plot shows the DN spikes.}
\label{fig:plots_AF}
\end{figure}

\ifdraft{\subsubsection*{LNA results}}

Figure~\ref{fig:plots_AF} shows experimental results measured from the different circuits present in the input headstage. 
Fig.~\ref{fig:plots_AF}a shows the transient response of the LNA to a prerecorded iEEG signal used as input. The signal was provided to the headstage directly via an arbitrary waveform generator programmed with unity gain and loaded with a sequence of the prerecorded iEEG data with amplitude in the mV range.
We also tested the LNA with an input sine wave of 100\,Hz with a 1\,mV peak-to-peak swing, revealing $<$1\% of total harmonic distortion at maximum gain, and an output swing ranging between 0.7\,V and 1.4\,V.
To characterize the input referred noise of the LNA, we shorted input terminals of the LNA, $V_{in+}, V_{in-}$, captured LNA output, $V_{amp}$, on a dynamic signal analyzer and divided it by the gain of the LNA, set to 100, and plotted output power spectral density (see Fig.~\ref{fig:plots_AF}b).
The LNA generates a $\le$ 100\,nV/$\sqrt{Hz}$ noise throughout the spectrum.
As the $1/f$ noise dissipates with the increase in frequency, the LNA noise only scales down to the thermal component. Thus,  the noise for the Ripple band is $<$  10\,nV/$\sqrt{Hz}$ and $<$ 5\,nV/$\sqrt{Hz}$ for the Fast Ripple band.
Figure~\ref{fig:plots_AF}b shows how the noise generated by the LNA is well below the pre-recorded iEEG power, throughout the full frequency spectrum.
The LNA features a programmable gain that can be set to 20\,dB, 32\,dB, 36\,dB, or 40\,dB. It has a $>$ 40\,dB common mode rejection ratio; 
it consumes 3\,$\mu$W of power per channel; and it has a total bandwidth, defined as $Gm$/($A_{M} \cdot C_L$), approximately equal to 11.1\,KHz, when the capacitive load is set to $C_L$=20\,fF, the OTA transconductance to $Gm$=20\,nS and the amplifier gain to 40\,dB (see Fig.~\ref{fig:plots_AF}c).

\ifdraft{\subsubsection*{Filter results}}

The on-chip bias generator can be used to set the filter frequency bands. For HFO detection, we biased the tunable pseudoresistors of the filters to achieve a cutoff frequency of 80\,Hz for the low-pass filter, a range between 80\,Hz and 250\,Hz for the first bandpass filter, appropriate for Ripple detection, and between 250\,Hz and 500\,Hz for the second bandpass filter, to detect Fast Ripples (see Fig.\ref{fig:plots_AF}c).
As we set the tail current of each single-stage OpAmp of Fig.~\ref{fig:schematic_AF}c to 150\,nA, each filter consumed 0.9\,$\mu$W of power.

\ifdraft{\subsubsection*{ADM results}}

Figure~\ref{fig:plots_AF}d plots the spikes produced by the AMD circuit in response to Ripple-band data obtained from the pre-recorded iEEG measurements. We set the ADM refractory period to 300\,$\mu$s, making it the longest delay, compared to those introduced by the comparator and handshaking circuits, that are typically $<$100\,$\mu$s. The spike-rate of the ADM can range from few hundreds of Hz to hundreds of kHz depending on the values of  $V_{tu}$, $V_{td}$, and $V_{refr}$. Each ADM consumed 1.5\,nJ of energy per spike, and had a static power dissipation of 96\,nW.
 
\begin{figure}
\centering\includegraphics[width=\linewidth]{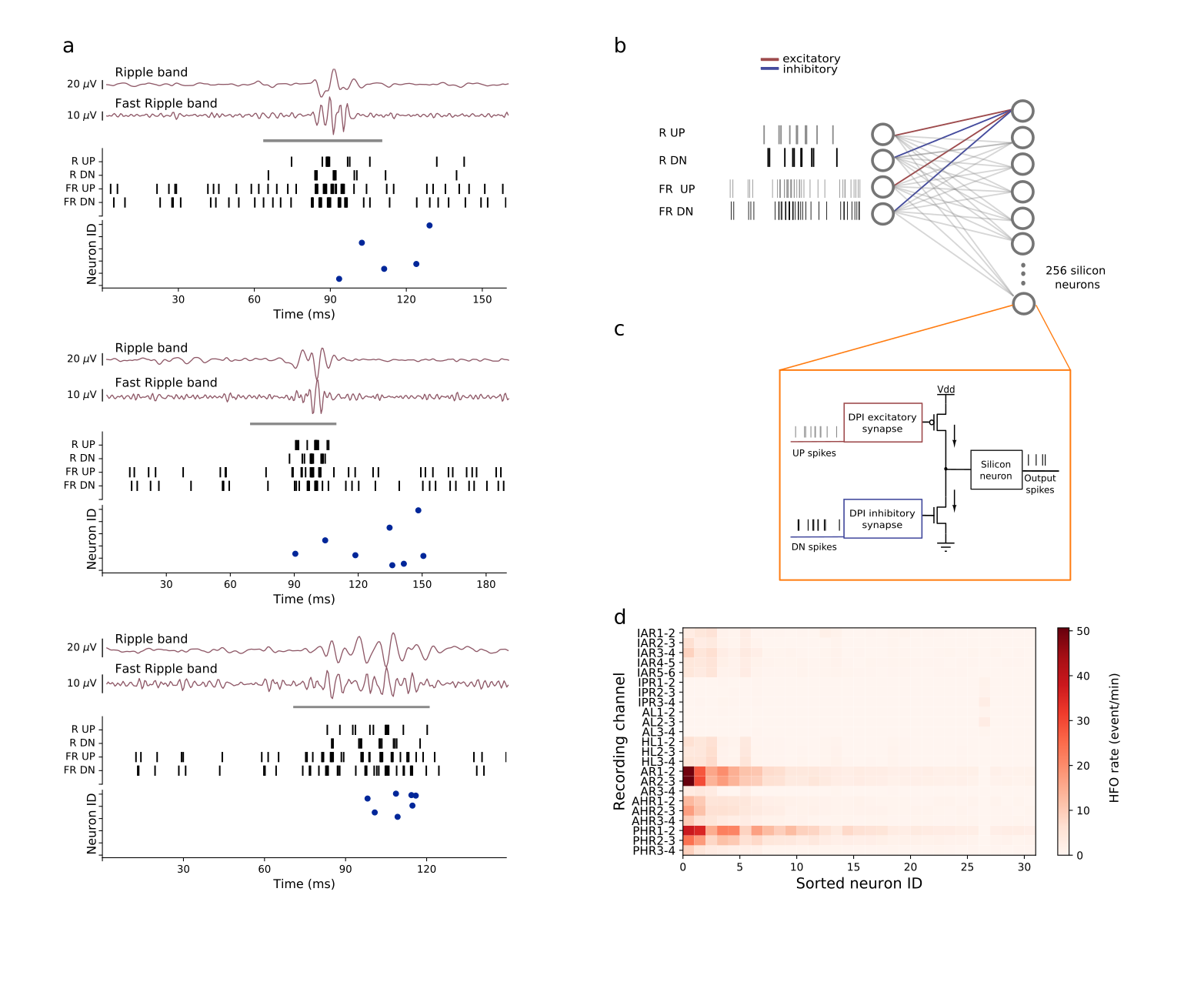}
  \caption{(a) Examples of HFOs
  that the hardware SNN detected in the iEEG of Patient 1.
  The periods marked by the gray bar represent clinically relevant HFO~\cite{Fedele_etal17b,Fedele_etal17c}.
  The signals in Ripple and Fast-ripple band were transformed to UP and DN spikes. These spike trains were sent to the neurons in the hardware through the $R_{UP}$, $R_{DN}$, $FR_{UP}$ and $FR_{DN}$ channels. The bottom panel of each example shows the raster plot of the silicon neurons. Each neuron responds to different HFOs depending on the characteristics of the pattern. (b) The SNN architecture consist of a two-layer network of 256 integrate and fire neurons with dynamic synapses. Each neuron in the second layer receives excitatory connections from the $R_{UP}$ and $FR_{UP}$ channels, and inhibitory connections from the $R_{DN}$ and $FR_{DN}$ channels. The synaptic parameters time constants and weights are distributed randomly within a predetermined optimal range. (c) Hardware building blocks used for the implementations of the SNN: the DPI synapse is a ``Differential-Pair Integrator'' circuit~\cite{Bartolozzi_Indiveri07a}, and the silicon neuron is an Adaptive Exponential Integrate and Fire (AdExp IF) circuit~\cite{Qiao_etal15}. (d) HFO rates computed for Patient 1. The neurons are sorted according to their average firing rate. Only a small number of neurons fire across all the recordings, even for channels with high HFO rates (e.g. AR1-2).}
 \label{fig:SNN_steps}
\end{figure}

\subsection{System performance}

The system-level performance is assessed by measuring the ability of the proposed device to correctly measure the iEEG signals, to properly encode them with spike trains, and to detect clinically relevant HFOs~\cite{Fedele_etal17b} via the SNN architecture.

\ifdraft{\subsubsection*{The SNN architecture}}

The SNN architecture is a two-layer feed-forward network of integrate and fire neurons with dynamic synapses, i.e., synapses that produce post-synaptic currents that decay over time with first-order dynamics.
The first layer of the network comprises four input neurons: the first neuron conveys the UP spikes that encode the iEEG signal filtered in Ripple band; the second neuron conveys the DN spikes derived from the same signal; the third neuron conveys the UP spikes derived from the Fast-ripple band signal, while the forth neuron conveys the DN spikes.
The second layer of the network contains 256 neurons which receive spikes from all neurons of the input layer (see Fig.~\ref{fig:SNN_steps}b).
The current produced by the dynamic synapses of the second layer neurons decay exponentially over time at a rate set by a synapse-specific time-constant parameter. The amplitude of these currents in response to the input spikes depends on a separate weight parameter, and their polarity (positive or negative) depends on the synapse type (excitatory or inhibitory).
In the network proposed, UP spikes are always sent to excitatory synapses and DN spikes to inhibitory ones. All neurons in the second layer have the same connectivity pattern as depicted in Fig.~\ref{fig:SNN_steps}b with homogeneous weight values.
An important aspect of the SNN network lies in the way it was configured to recognize the desired input spatio-temporal patterns: rather than following the classical Artificial Neural Network (ANN) approach of training the network by modifying the synaptic weights with a learning algorithm and using identical settings for all other parameters of synapses and neurons, we fixed the weights to constant values and chose appropriate sets of parameter distributions for the synaptic time constants and neuron leak variables.
Because of the different time-constants for synapses and neurons, the neurons of the second layer produce different outputs, even though they all receive the same input spike trains.

\ifdraft{\subsubsection*{Software simulations} }

The set of parameter distributions that maximized the network's HFO detection abilities was found heuristically by analyzing the temporal characteristics of the input spike trains and choosing the relevant range of excitatory and inhibitory synapse time constants that produced spikes in the second layer only for the input signals that contained HFOs as marked by the Morphology Detector~\cite{Fedele_etal17b} 
(Figure 1a, Figure 5a). 
This procedure was first done using software simulations with random number generators and then validated in the neuromorphic analog circuits, exploiting their device mismatch properties.
The software simulations were carried out using a behavioral-level simulation toolbox~\cite{Milde_etal18} based on the neuromorphic circuit equations, that accounts for the properties of the mixed-signal circuits in the hardware SNN.
\ifdraft{\subsubsection*{Hardware implementation of the SNN}}
The hardware validation of the network was done using a single core of the DYNAP-SE neuromorphic processor~\cite{Moradi_etal18}, which is a previous generation chip functionally equivalent to the one proposed, implemented using the same CMOS 180\,nm technology node. The 256 neurons in this core received spikes produced by the ADM circuits of the analog headstage, as described in Fig.~\ref{fig:SNN_steps}c. The ADM UP spikes were sent to the excitatory synapses, implemented using a DPI circuit~\cite{Chicca_etal14,Milde_etal18} that produce positive currents, and DN spikes were sent to complementary versions of the circuit that produce inhibitory synaptic currents. Both excitatory and inhibitory currents were summed into the input nodes of their afferent leaky integrate-and-fire silicon neuron circuit~\cite{Chicca_etal14}, which produced output spikes only if both the frequency and the timing of the input spikes was appropriate (see Fig.~\ref{fig:SNN_steps}c).
The bias values of the excitatory and inhibitory DPI circuits and of the neuron leak circuits were set in a way to match the mean values of the software simulation parameters. All neuron and synapse circuits in the same core of the chip share global bias parameters, so nominally all excitatory synapses would have the same time-constant, all inhibitory ones would share a common inhibitory time-constant value and all neurons would share the same leak parameter value. However, as the mixed-signal analog-digital circuits that implement them are affected by device mismatch, they exhibit naturally a diversity of behaviors that implements the desired variability of responses. Therefore, in the hardware implementation of the SNN, the distribution of parameters that produce the desired different behaviors in the second layer neurons emerges naturally, by harnessing the device mismatch effects of the analog circuits used and without having to use dedicated random number generators~\cite{Payvand_etal19}.

\ifdraft{\subsubsection*{Hardware HFO detection}}

Figure~\ref{fig:pipeline_combined}e (bottom panel) shows an example of the activity of the hardware SNN in response to an HFO that was labeled as clinically relevant by the Morphology Detector~\cite{Fedele_etal17b} 
The iEEG traces in the Ripple band and Fast ripple band (Fig.~\ref{fig:pipeline_combined}a) and the time frequency spectrum (top panel of Fig.~\ref{fig:pipeline_combined}e) show the HFO shortly before the SNN neurons spike in response to it (bottom panel of Fig.~\ref{fig:pipeline_combined}e).
The delay between the beginning of the HFO and the spiking response of the silicon neurons is due to the integration time of both excitatory and inhibitory synapse circuits, which need to accumulate enough evidence for producing enough positive current to trigger the neuron to spike. 

To improve classification accuracy and robustness, we adopted an \emph{ensemble} technique~\cite{Polikar06} by considering the assembly response of the 256 neurons in the network: the system is said to detect an HFO if at least one neuron in the second layer spikes in a 15\,ms interval.
We counted the number of HFOs detected per electrode channel and computed the corresponding HFO rate (Section~\ref{sec:methods}).
Examples of HFOs recorded from Patient 1 and detected by the hardware SNN are shown in Fig.~\ref{fig:SNN_steps}a; several neurons respond within a few milliseconds after initiation of the HFO. Different HFOs produce different UP and DN spike trains, which in turn lead to different sets of second layer neurons spiking. Figure~\ref{fig:SNN_steps}d shows the HFO rates calculated for each electrode from the recordings of this patient. Observe that not all the neurons in the second layer respond to the HFOs. Even for electrode channels with high HFO rates, a very small number of neurons fire at high rates.

The robustness of the HFO rate measured with our system can be observed in Fig~\ref{fig:pipeline_combined}i, where the relative differences of HFO rates across channels in Patient 1 persisted over multiple nights. To quantify this result we performed a test-retest reliability analysis by computing the scalar product of the HFO rates across all recording intervals (0.95 in Patient 1), where the scalar product is \~1 for highly overlapping spatial distributions, indicating that the HFO distribution persists over intervals.

\subsection{Predicting seizure outcome}

\newcommand{\ra}[1]{\renewcommand{\arraystretch}{#1}}
\begin{table}
\ra{1.3}
\caption{ Patient characteristics and postsurgical seizure outcome. We 'predict' seizure outcome for each patient based on resection of the HFO area that was delineated by the Morphology Detector~\cite{Fedele_etal17b} and the hardware SNN of our system.
HS hippocampal sclerosis; ILAE seizure outcome classification of the International League Against Epilepsy.}
\label{table:patient_characteristics}
\renewcommand\arraystretch{1.2}
\centering
\resizebox{\textwidth}{!}{\begin{tabular}{@{}cccccccc@{}}
\toprule
\textbf{Patient} & \textbf{Histology/} & \textbf{Intervals} & \textbf{Test-retest} &  \textbf{Outcome}& \textbf{Follow-up} &\textbf{Morphology Detector} & \textbf{Hardware SNN}\\
 &  \textbf{Pathology} & \textbf{of 5 min} &\textbf{intervals} & \textbf{(ILAE)} &\textbf{(months)} &\textbf{prediction}  & \textbf{prediction}\\
\hline
1 & HS  &28 & 0.95 & 1 & 12 & TN  & TN\\
2 & Glioma &13 & 0.97 & 1 & 29 & TN  & TN\\
3 & HS  &39 & 0.83 & 1 & 13 & TN  & TN\\
4 & HS  &34 & 0.96 & 1 & 41 & TN  & TN\\
5 & HS  &35 & 0.91 & 1 & 14 & TN  & TN\\
6 & HS  &35 & 0.59 & 1 & 11 & TN  & TN\\
7 & HS  &1 &   --  & 3 & 42 & FN  & FN\\
8 & HS  &16 & 0.74 & 3 & 15 & FN  & TP\\
9 & HS  &12 & 0.90 & 5 & 46 & FN  & FN\\
\bottomrule
\end{tabular}}
\vspace{-0.31cm}
\end{table}

In Patient 1, the electrodes were implanted in right frontal cortex (IAR, IPR), the left medial temporal lobe (AL, HL) and the right medial temporal lobe (AR, AHR, PHR). The recording channels AR1-2 and AR2-3 in the right amygdala produced the highest HFO rates persistently. We included all channels with persistently high HFO rate in the 95\% percentile to define the ``HFO area''. If the HFO area is included in the resection area, resection area (RA), we would retrospectively `predict' for the patient to achieve seizure freedom. Indeed, right selective amygdalohippocampectomy in this patient achieved seizure freedom for >1 year.  

We validated the system performance across the whole patient group by performing the test-retest reliability analysis of all the data. 
The test-retest reliability score ranges from 0.59 to 0.97 with a median value of 0.91. 
We compared the HFO area detected by our system with the RA.
For each individual, we then retrospectively determined whether resection of the HFO area would have correctly `predicted’ the postsurgical seizure outcome (Table~\ref{table:patient_characteristics}). 
Seizure freedom (ILAE 1) was achieved in 6 of the 9 patients. 
To estimate the quality of our `prediction', we classified each patient as follows: we defined as ``True Negative'' (TN) a patient where the HFO area was fully located inside the RA and who became seizure free; ``True Positive'' (TP) a patient where the HFO area was not fully located within the RA and the patient suffered from recurrent seizures; ``False Negative'' (FN) a patient where the HFO area was fully located within the RA but who suffered from recurrent seizures; ``False Positive'' (FP) a patient where the HFO area was not fully located inside the RA but who nevertheless achieved seizure freedom.

The HFO area defined by our system was fully included in the RA in patients 1 to 6. These patients achieved seizure freedom and were therefore classified as TN. In Patients 7 and 9, the HFO area was also included in the RA but these were classified as FN since these patients did not achieved seizure freedom. In Patient 8, the HFO area was not included in the RA and the patient did not achieve seizure freedom (TP). 

We finally compared the predictive power of our detector to that of the Morphology Detector~\cite{Fedele_etal17b} for the individual patients (Table~\ref{table:patient_characteristics}) and over the group of patients (Table~\ref{tab:metrics}).
The overall prediction accuracy of our system across the 9 patients is comparable to that obtained by the Morphology Detector. 
The 100\% specificity achieved by both detectors indicates that HFO analysis provides results consistent with the current surgical planning.

\begin{table}
\ra{1.3}
	\caption{Comparison of postsurgical outcome prediction between the Morphology Detector and our system. TP True Positive; TN True Negative; FP False Positive; FN False Negative; N = TP + TN + FP + FN = number of patients. The Morphology Detector did not classify a TP so that sensitivity and PPV can not be calculated.}
	\label{tab:metrics}
	\renewcommand\arraystretch{1.2}
	\centering
	\begin{tabular}{@{}l >{\centering\arraybackslash}p{0.15\textwidth} >{\centering\arraybackslash}p{0.2\textwidth} >{\centering\arraybackslash}p{0.2\textwidth} @{}}
		\toprule
		&\textbf{Morphology Detector prediction [\%]} &\textbf{Hardware SNN prediction [\%]}\\
		\hline
		\textbf{Specificity} = $TN/(TN + FP)$  & 100  & 100\\
		\textbf{Sensitivity} = $TP/(TP + FN)$  & -- & 33\\
		\textbf{Negative Predictive Value} = $TN/(TN + FN)$ &67 &75\\
		\textbf{Positive Predictive Value} = $TP/(TP + FP)$& --&100\\ 
		\textbf{Accuracy} = $(TP + TN)/N$ [\%]& 67 & 78\\
		\bottomrule
	\end{tabular}
	\vspace{-0.31cm}
\end{table}

\section{Discussion}

The device presented here demonstrates the potential of neuromorphic computing for extreme-edge use
cases that cannot rely on computation performed on remote computers (e.g., in the ``cloud''), and that require compact and very low power embedded systems (e.g., for battery-operated hand-held operations). By integrating on the same chip both the signal acquisition headstage and the neuromorphic multi-core processor, we developed an integrated system that can demonstrate the advantages of neuromorphic computing in clinically relevant applications. 

Although numerous neural recording headstages have been already developed and optimized for their specific application domain (e.g., for very large scale arrays~\cite{Jun_etal17,Frey_etal10,Ballini_etal14,Khazaei_etal20}, or for intracranial recordings\cite{Harrison_etal07,Wattanapanitch_etal07, Angotzi_etal19, Kim_etal19, Wang_etal19, Valtierra_etal20, Ture_etal20}),
to our knowledge, this is the first instance of a headstage design that has the capability of adapting to numerous use cases requiring different gain factors and band selections, on the same input channel. 

The approach followed to process the signals in the SNN and determine the right architecture for detecting HFOs is radically different from the deep-learning one: rather than using arrays of neurons with homogeneous parameters and no temporal dynamics, and relying on the changes in the synaptic weights, we emulated the dynamic properties of real neurons with mixed-signal analog/digital circuits and exploited their variability to find the right set of randomly distributed parameters for tuning the network to detect HFOs.

\ifdraft{\subsection*{Discussion on SNN HFO detector}}

The simulations of the SNN not only allowed us to define the optimal architecture for HFO detection, but also gave us solutions for setting the hyperparameters of the analog headstage, such as the refractory period $V_{ref}$ and the threshold ($V_{tu}$ and $V_{td}$) for the signal-to-spike conversion of the ADM . 
While mismatch effect is generally a concern in modeling hardware designs based on software simulations, we show here that the mismatch among the silicon neurons resulted in a key feature for the implementation of our SNN architecture. This advantage allowed us to generate the normal distribution of parameters without manually defining the distribution of neuronal time-constants found in simulations or requiring extra memory to allocate these values.

\ifdraft{\subsubsection*{HFO detected by a SNN predict seizure freedom in the individual patient}}

By averaging over both time and the number of neurons recruited by the \emph{ensemble} technique, the SNN network was able to achieve robust results: the accuracy obtained by the SNN are compatible with those obtained by state-of-the-art software algorithms implemented using complex algorithms on powerful computers~\cite{Fedele_etal17b}.
Overall, the high specificity (100\%) achieved with our system not only generalizes the value of the detected HFO by the SNN across different types of patients, but still holds true at the level of the individual patients,  which is a prerequisite to guide epilepsy surgery that aims for seizure freedom.

\section{Conclusions}

This is the first feasibility study towards identifying relevant features in intracranial human data in real-time, on-chip, using event-based processors and spiking neural networks. By providing ``neuromorphic intelligence'' to neural recording circuits the approach proposed will lead to the development of systems that can detect HFO areas directly in the operation room and improve the seizure outcome of epilepsy surgery.

\section{Methods}
\label{sec:methods}

\subsection{Design and setup of the hardware device}

The CMOS circuit simulations were carried-out using the Cadence\textsuperscript\textregistered{} Virtuoso ADE XL design tools. All circuits including the headstage, the bias generator, and the silicon neurons were designed, simulated and analyzed in analog domain. The asynchronous buffers, spike routing network and chip configuration blocks were simulated and implemented in the asynchronous digital domain. The layout of the chip was designed using the Cadence\textsuperscript\textregistered{} Layout XL tool. The design rule check, layout versus schematic and post-layout extraction were performed using the Calibre tool.
We packaged our device using a ceramic 240-pin quadratic flat package. The package was then mounted on an in-house designed six-layer printed circuit board. The programming and debugging of the System-on-Chip (SoC) was performed using low-level software and firmware developed in collaboration with SynSense Switzerland, and implemented using the XEM7360 FPGA (Opal Kelley, USA). The pre-recorded iEEG was fed to the chip using a Picoscope 2205A MSO (Picotech, UK). All frequency-domain measurements were performed using a Hewlett-Pacard 35670A dynamic signal analyzer.

\subsection{Simulation and validation of the neural architecture}
\label{sec:sw_methods} 

For the software simulation of the network we used the Spiking Neural Network simulator Brian2~\cite{Goodman_Brette08} and a custom made toolbox~\cite{Milde_etal18} that makes use of equations which describe the behavior of the neuromorphic circuits. To find the optimal parameters of the SNN, we were guided by the clinically relevant HFO marked by the Morphology Detector~\cite{Fedele_etal17b}: Around the HFOs marked in the iEEG~\cite{Fedele_etal17b} we created snippets of data $\pm$25 ms. These snippets were used to select the parameters for the ADM and the SNN network (see Methods).
The SNN architecture was validated using the previous generation of the neuromorphic processor DYNAP-SE~\cite{Moradi_etal18}, for which a working prototyping framework is available. The high-level software-hardware interface used to send signals to the SNN, configure its parameters, and measure its output was designed in collaboration with SynSense AG, Switzerland.


\subsection{Patient data}
\label{sec:data}

We included long-term iEEG recordings from 9 patients of a data set that is publicly available~\cite{Fedele_etal17c}. Patients had drug-resistant focal epilepsy as detailed in Table~\ref{table:patient_characteristics}.
Presurgical diagnostic workup at Schweizerische Epilepsie-Klinik included recording of iEEG from the medial temporal lobe.
The iEEG data set has been analysed for HFO with state-of-the-art HFO detectors~\cite{Fedele_etal17b, Farahmand_etal19}.

The surgical planning technique~\cite{Zijlmans_etal19} was independent of HFO.
Patients underwent resective epilepsy surgery at UniversitätsSpital Zürich.
After surgery, the patients were followed-up for >1 year.
Postsurgical outcome was classified according to the International League Against Epilepsy (ILAE):
\begin{description*}
\item[Class 1] Completely seizure free; no auras. 
\item[Class 2] Only auras; no other seizures.
\item[Class 3] One to three seizure days per year; $\pm$ auras.
\item[Class 4] Four seizure days per year to 50\% reduction of baseline seizure days; $\pm$ auras.
\item[Class 5] Less than 50\% reduction of baseline seizure days to 100\% increase of baseline seizure days; $\pm$ auras.
\item[Class 6] More than 100\% increase of baseline seizure days.
\end{description*}

\ifdraft{\subsubsection*{Previous HFO analysis of the same data with the Morphology Detector}}

In a previous publication~\cite{Fedele_etal17b}, we detected HFOs with the Morphology detector~\cite{Fedele_etal17b, Burnos_etal16}
and compared the HFO area to the resected area to predict seizure outcome. 
The dataset with the HFO markings~\cite{Fedele_etal17c} is publicly available and was downloaded for the analyses presented here.
The data consist of up to six intervals of approximately five minutes that were recorded in the same night during interictal slow-wave sleep, which promotes low muscle activity and high HFO rates.
The intervals were at least three hours apart from epileptic seizures to eliminate the influence of seizure activity on the analysis. The amount of nights and intervals varied across patients (see Table~\ref{table:patient_characteristics}). 
Because of the higher signal-to-noise ratio, in this study we focused on the recordings from the medial temporal lobe.
Only the 3 most mesial bipolar channels were included in the analysis. 

\subsection{HFO detection}
\label{sec:hfo_detection}

HFO detection was performed independently for each channel in each 5-min interval of iEEG.
The signal pre-processing steps consisted of bandpass filtering, baseline detection and transforming the continuous signal into spikes using the ADM block. 
The ADM principle of operation is as follows: whenever the amplitude variation of the input waveform exceeds an upper threshold $V_{tu}$ a positive spike on the UP channel is generated; if the change in the amplitude is lower than a threshold $V_{td}$, a negative spike in the DN channel is produced.

\ifdraft{\subsubsection*{Baseline detection}}

As the amplitude of the recordings changed dramatically with electrode, patient data and recording session, we introduced a baseline detection mechanisms that was used to adapt the values of the $V_{tu}$ and $V_{td}$ thresholds in order to produce the optimal number of spikes required for detecting HFO signals while suppressing the background noise and outliers in the recordings (see Fig.~\ref{fig:baseline}).
This baseline was calculated for each iEEG channel in software: during the first second of recording, the maximum signal amplitude was computed over non-overlapping windows of 50\,ms. These values were then sorted and the baseline value was set to the average of the lowest quartile. This procedure excluded outliers on one hand, and suppressed the noise floor on the other hand.
This procedure was optimal for converting the recorded signals into spikes.

\begin{figure}
  \centering
  \includegraphics[width=0.75\textwidth]{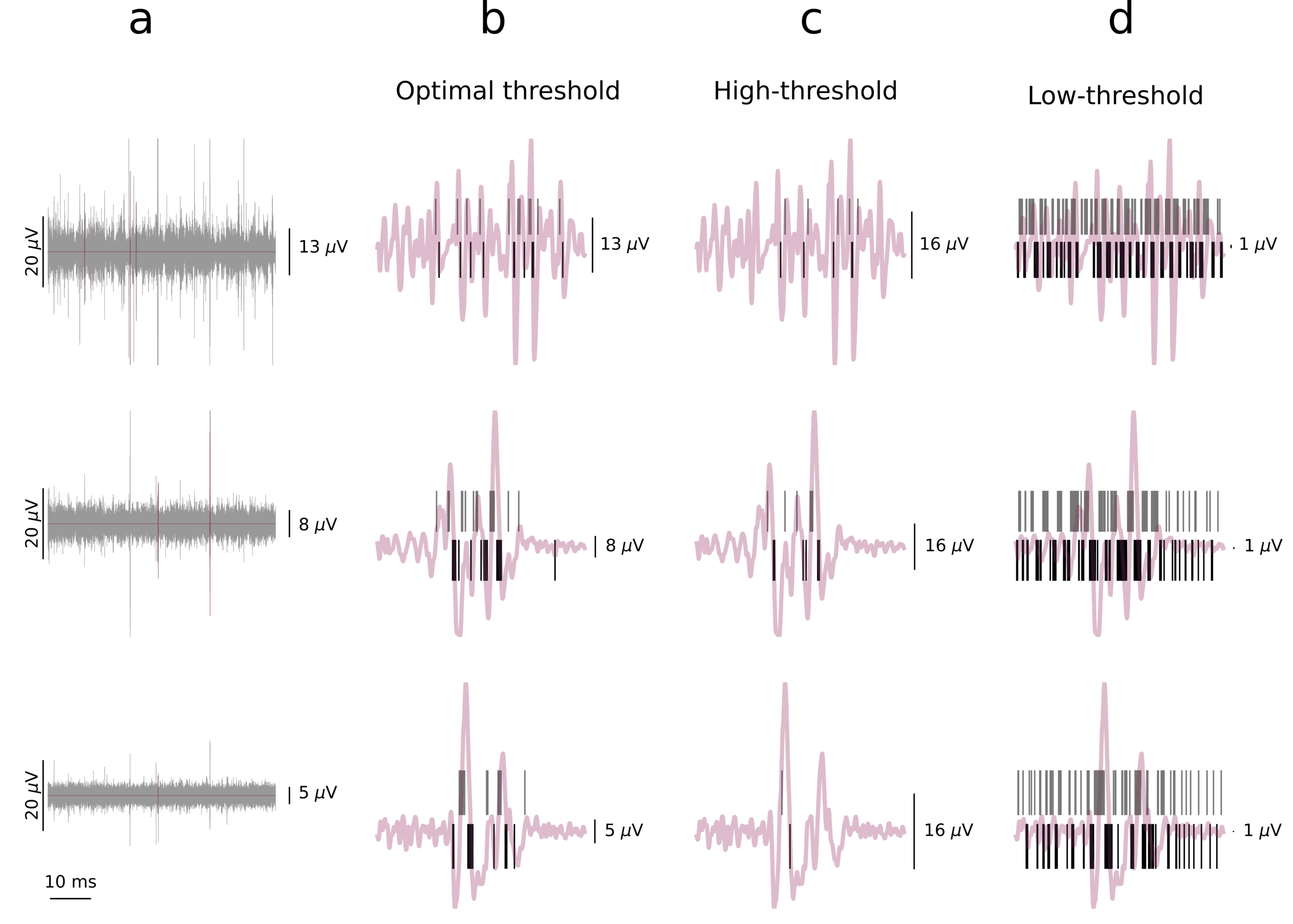}
  \caption{Optimal spike generation threshold for the asynchronous delta modulator (ADM). 
  (a) iEEG from 3 channels in one electrode.
  The different noise floors are captured by the different baseline levels of [5, 8, 13] $\mu$V that we computed. 
  (b) With the optimal channel-wise threshold, the number of UP and DN spikes is optimal for the example HFOs taken from each channel. 
  (c) With a higher threshold (16 $\mu$V), the number of UP and DN spikes is too low for the example HFOs.
  (d) With a lower threshold (1 $\mu$V), the number of UP and DN spikes is too high for the example HFOs.}
  \label{fig:baseline}
\end{figure}

\ifdraft{\subsubsection*{SNN parameters}}

 Spikes entered the SNN architecture as depicted in Fig.~\ref{fig:SNN_steps}bc.
The SNN parameters used to maximize HFO detection were selected by analyzing the Inter-Spike-Intervals (ISIs) of the spike trains produced by the ADM and comparing their characteristics in response to inputs that included an HFO event versus inputs that had no HFO events.
This analysis was then used to tune the time constants of the SNN output layer neurons and synapses.
Specifically, the approach used was to rely on an \emph{ensemble} of neurons in the output layer and to tune them with parameters sampled from a uniform distribution. The average time constant for the neurons was chosen to be 15\,ms, with a coefficient of variation set by the analog circuit devise mismatch characteristics, to approximately 20\%. Similarly, the excitatory synapse time constants were set in the range [3 6]\,ms and the inhibitory synapse time constants in the range [0.1 1]\,ms.

After sending the spikes produced by the ADMs to the SNN configured in this way, we evaluated snippets of 15 milliseconds output data produced by the SNN and signaled the detection of an HFO every time spikes were present in consecutive snippets of data.
Outlier neurons in the hardware SNN that spiked continuously were considered uninformative and were switched off for the whole study.
The activity of the rest of the neurons faithfully signaled the detection of HFOs (see Table~{tab:metrics}). For the HFO count, spikes with inter-spike-intervals <15 ms were aggregated to mark a single HFO.


\subsection{Post-surgical outcome prediction}
\label{sec:delineating the hfo_area}

To retrospectively 'predict' the postsurgical outcome of each patient in this data set, we first detected the HFOs in each 5-min interval by measuring the activity of the silicon neurons in the hardware SNN.
We calculated the rate of HFO per recording channel by dividing the number of HFOs in the specific channel by the duration of the interval.
The distribution of HFO rates over the list of channel defines the HFO vector.
In this way we calculated an HFO vector for each interval in each night. We quantified the test-retest reliability of the distribution of HFO rates over intervals by computing the scalar product of all pairs of HFO vectors across intervals (Table~\ref{table:patient_characteristics}).
We then delineated the ``HFO area'' by comparing the average HFO rate over all recordings and choosing the area under the electrodes with HFO rates exceeding the 95 percentile of the rate distribution. 
Finally, to assess the accuracy of the patient outcome prediction, we compared the HFO area identified by our procedure with the area that was resected in surgery, and compared it with the postsurgical seizure outcome (Table~\ref{table:patient_characteristics}).

\section*{Acknowledgements}
This project has received funding from Swiss National Science Foundation (SNSF 320030$\_$176222) and from the European Research Council (ERC) under the European Union’s Horizon 2020 research and innovation program grant agreement No 724295.

\section*{Author contributions statement}

JS and GI conceived the experiments, MS and KB conducted the experiments and analysed the results. All authors wrote and reviewed the manuscript. 




\bibliography{biblioncs}

\end{document}